\documentclass[12pt]{article}
\usepackage{amssymb,amsmath,amsfonts,amsthm,amstext,amscd,array,cite}
\usepackage{mathrsfs}
\usepackage{hyperref}
\usepackage{pdfsync}
\usepackage{bbm}
\usepackage{bm}
\usepackage[arrow,matrix,curve]{xy}
\usepackage{bbding}
\usepackage{wasysym}

\usepackage{amsfonts}
\usepackage{booktabs}
\usepackage{siunitx}

\usepackage[utf8]{inputenc}
\usepackage{amsthm,amsmath,latexsym,amssymb,amsfonts,amscd,cite}
\usepackage{amssymb} 

\usepackage{tocbasic}
\DeclareTOCStyleEntry[
  beforeskip=.2em plus 1pt,
  pagenumberformat=\textbf
]{tocline}{section}

\usepackage{mathrsfs}
\usepackage{hyperref}

\usepackage{epigraph}

\usepackage{bbm}
\usepackage{bm}
\usepackage[T1]{fontenc}
\usepackage{mathrsfs}

\usepackage{bbm}
\usepackage{bm}
\usepackage[T1]{fontenc}
\usepackage{mathrsfs}

\usepackage{epsf}
\usepackage{epsfig}
\usepackage{wrapfig}
\usepackage[utf8]{inputenc}

\usepackage{mathrsfs}
\usepackage{hyperref}

\usepackage{amsthm,amsmath,latexsym,amssymb,amsfonts,amscd}
\usepackage{authblk}

\input{epsf}

\def\dd{{\rm d}}

\def\sfs{{\sf s}}
\def\sft{{\sf t}}
\def\sfi{{\sf i}}

\newcommand{\eq}{\begin{equation}}
\newcommand{\eqend}{\end{equation}}
\newcommand{\eqa}{\begin{eqnarray}}
\newcommand{\nonueqa}{\begin{eqnarray*}}
\newcommand{\eqaend}{\end{eqnarray}}
\newcommand{\nonueqaend}{\end{eqnarray*}}

\newcommand{\bma}[1]{\begin{array}{#1}}
\newcommand{\ema}{\end{array}}
\newcommand{\bc}{\begin{center}}
\newcommand{\ec}{\end{center}}

\newif\ifold             \oldtrue

\def\be{\begin{equation}}
\def\ee{\end{equation}}
\def\bea{\begin{eqnarray}}
\def\eea{\end{eqnarray}}
\def\bd{\begin{displaymath}}
\def\ed{\end{displaymath}}

\newcommand{\hh}{ \mathbb{H}}

\newcommand{\beq}{\begin{eqnarray}}
\newcommand{\eeq}{\end{eqnarray}}

\makeatletter
\newdimen\normalarrayskip              
\newdimen\minarrayskip                 
\normalarrayskip\baselineskip
\minarrayskip\jot
\newif\ifold             \oldtrue            
\def\arraymode{\ifold\relax\else\displaystyle\fi} 
\def\@arrayskip{\ifold\baselineskip\z@\lineskip\z@
     \else
     \baselineskip\minarrayskip\lineskip2\minarrayskip\fi}
\def\@arrayclassz{\ifcase \@lastchclass \@acolampacol \or
\@ampacol \or \or \or \@addamp \or
   \@acolampacol \or \@firstampfalse \@acol \fi
\edef\@preamble{\@preamble
  \ifcase \@chnum
     \hfil$\relax\arraymode\@sharp$\hfil
     \or $\relax\arraymode\@sharp$\hfil
     \or \hfil$\relax\arraymode\@sharp$\fi}}
\def\@array[#1]#2{\setbox\@arstrutbox=\hbox{\vrule
     height\arraystretch \ht\strutbox
     depth\arraystretch \dp\strutbox
     width\z@}\@mkpream{#2}\edef\@preamble{\halign \noexpand\@halignto
\bgroup \tabskip\z@ \@arstrut \@preamble \tabskip\z@ \cr}%
\let\@startpbox\@@startpbox \let\@endpbox\@@endpbox
  \if #1t\vtop \else \if#1b\vbox \else \vcenter \fi\fi
  \bgroup \let\par\relax
  \let\@sharp##\let\protect\relax
  \@arrayskip\@preamble}
\makeatother

\def\be{\beta}

\newtheorem{definition}[equation]{Definition}

\theoremstyle{definition}
\newtheorem{remark}[equation]{Remark}

\usepackage{tempora}
\usepackage{color,xcolor}
\usepackage{manfnt}

\def\ddo{\end{document}}
\usepackage{hyperref}
\hypersetup{colorlinks,%
citecolor=black,%
filecolor=black,%
linkcolor=black,%
urlcolor=black}

\usepackage{tocbasic}
\DeclareTOCStyleEntry[
  beforeskip=.2em plus 1pt,
  pagenumberformat=\textbf
]{tocline}{section}

\usepackage{mathrsfs}
\usepackage{hyperref}

\usepackage{bbm}
\usepackage{bm}
\usepackage[T1]{fontenc}
\usepackage{mathrsfs}

\title{\bf  Classical Mechanics in Noncommutative Spaces:\\Confinement and More }

\author[1]{Vladislav \textsc{Kupriyanov}} 
\author[2]{Maxim \textsc{Kurkov}}
\author[3]{Alexey \textsc{Sharapov}}
\affil[1] {\it \small Centro de Matem\'atica, Computa\c{c}\~{a}o e
Cogni\c{c}\~{a}o, Universidade Federal do ABC, Santo Andr\'e, SP, 
Brazil} 
\affil[2]{\it \small Dipartimento di Fisica “E. Pancini”, Universit\`a di Napoli Federico II, Complesso Universitario di Monte S. Angelo Edificio 6, via Cintia, 80126 Napoli, Italy.}
\affil[3]{\it \small Physics Faculty, Tomsk State University, Lenin ave. 36, Tomsk 634050, Russia}
\date{}

\begin{document}

\maketitle

\begin{abstract}
We consider a semi-classical approximation to the dynamics of a point particle in a noncommutative space. In this approximation, the noncommutativity of space coordinates is described by a Poisson bracket. 
For linear Poisson brackets, the corresponding phase space is given by the
cotangent bundle of a Lie group, with the Lie group playing the role of a curved momentum space. We show that the curvature of the momentum space may lead to rather unexpected physical phenomena such as an upper bound on the velocity of a free nonrelativistic particle,  bounded motion for repulsive central force, and no-fall-into-the-centre for attractive Coulomb potential.  We also consider a superintegrable Hamiltonian for the Kepler problem in $3$-space with $\mathfrak{su}(2)$ noncommutativity.  The leading correction to the equations of motion due to noncommutativity is shown to be described by an effective monopole potential.

\end{abstract}

\section{Introduction}
The idea of noncommutative spacetime, being quite natural, goes back to the early days of quantum theory. 
Indeed, if position and momentum coordinates become noncommuting operators when quantized, why not assume the same for position coordinates alone?
Then the generalized uncertainty principle $\Delta x^a\Delta x^b\geq \ell^2$ would bring about a fundamental length $\ell$, setting the absolute limit 
for locating a spacetime point. Intuitively, abandoning local interactions of particles could solve, or at least alleviate, the problem of ultraviolet divergences. This line of reasoning was first spelt out by Snyder \cite{Snyder1}. Further developments in quantum field theory, particularly the success of the renormalization program in quantum electrodynamics, completely eclipsed these ideas for many decades. Renewed interest in noncommutative spaces was sparked by the work of Seiberg and Witten \cite{Seiberg_1999} on the eve of the millennium. They considered the quantization of open strings ending on $D$-branes in the presence of a constant $B$-field. It was shown that the low-energy effective field theory on a single $D$-brane has a consistent deformation to a noncommutative gauge theory. The subsequent explosion of papers on noncommutative field theory (see \cite{Douglas:2001ba, Szabo:2001kg, Aschieri_2023,Hersent:2022gry} for a comprehensive review) was also supported by parallel advances in mathematics, most notably the theory of deformation quantization, quantum groups, and noncommutative geometry. 

Another old proposal for eliminating ultraviolet divergences is based on the hypothesis of curved momentum space. This was first put forward by Born \cite{Born, born1949reciprocity} and then considered by many authors (see e.g. \cite{Golfand:1962kjf, Tamm, Kadyshevsky:1977mu,Freidel:2005me,Amelino, KOWALSKI_GLIKMAN_2013,Guedes:2013vi,Smilga:2022xij,Franchino-Vinas:2023rcc, A5} and references therein). 
 If the particle's momentum space is `curved enough' to have a finite volume, then we get an effective cut-off that makes all loop integrals finite at the upper limit. In particular, one can consider a compact momentum space, in which case there exists an upper bound on the uncertainty of momenta,  $\Delta p\leq M$. According to Heisenberg's uncertainty relation $\Delta x\Delta p\geq \hbar/2$, this yields immediately a lower bound for the uncertainty of positions $\Delta x\geq \hbar/M$. Thus, both the hypotheses -- noncommutativity and curvature -- introduce a fundamental length scale $\ell=\hbar/M$, leading to the granular structure of spacetime. This also indicates a certain connection between the curvature of momentum space and the noncommutativity of spacetime.

Since Born's pioneering paper \cite{Born}, the curvature of the momentum space has usually been treated within the framework of Riemannian  geometry. In our opinion, this choice of geometric structure was motivated mainly by a formal analogy with general relativity, without any deep physical or mathematical reasons. It is also clear that there is no point in considering momentum space separately from configuration one since only together do they constitute the phase space of a scalar particle. If the configuration space turns out to be noncommutative, this should affect the corresponding momentum space and vice versa. In the recent paper \cite{kupriyanov2024symplectic}, this interplay between the curvature of momentum space and the noncommutativity of spacetime  was formalized mathematically in the framework of symplectic groupoids. The idea was to consider semi-classical approximation wherein the noncommutativity of spacetime coordinates is controlled by their Poisson brackets. Upon (strict, deformation, geometric, canonical, etc.) quantization, these Poisson brackets are expected to give rise to a noncommutative algebra of quantum observables generated by the operators of spacetime coordinates.   
The whole phase space of a scalar particle is then identified with a {\it symplectic groupoid} integrating the underlying Poisson bracket.
Such integration is not always possible, but when it is possible, it is essentially unique. 
For example, the Lie-algebraic noncommutativity defines and is defined by linear Poisson brackets of coordinates, 
$\{x^a,x^b\}=f^{ab}_cx^c$. In this case, the conjugated momentum space is given by a Lie group integrating the corresponding Lie algebra, and the total phase space is the cotangent bundle of the Lie group endowed with the canonical symplectic structure. As is well known, there is a unique simply connected Lie group with a given Lie algebra. If the Lie algebra happens to be compact, then so is the corresponding Lie group. This provides a large number of examples of mechanical systems with compact momentum spaces. For a simple Lie algebra,  the corresponding Lie group enjoys a metric that is invariant under the right and left translations. This metric, being unique up to a constant factor, allows us to speak about the momentum space as a (pseudo-)Riemannian manifold. 
It should be emphasized that no canonical metrics exist on momentum spaces associated with general (linear or nonlinear) Poisson brackets. From this perspective, the existence of (pseudo-)Riemannian metrics for simple Lie groups and algebras is an accident. 

Not only  does the geometry of symplectic groupoids underlay the kinematics of point particles, but it also provides a geometric framework for formulating Poisson electrodynamics  \cite{Kupriyanov_2021,Kupriyanov:2021aet,Kupriyanov:2022ohu}, which is defined as a semi-classical limit of noncommutative gauge theories.  As explained in \cite{kupriyanov2024symplectic}, the gauge fields enter this picture as {\it bisections} of a symplectic groupoid.  The bisections form an infinite-dimensional Lie group and describe the low-energy limit of the gauge vector field in noncommutative spacetime.  Unlike Maxwell's electrodynamics, neither the target space nor the field equations of Poisson electrodynamics are linear. This nonlinearity reflects the noncommutativity of spacetime and disappears in the commutative limit. The conventional $U(1)$ gauge group is replaced here by the subgroup of Lagrangian bisections, which act on the gauge fields by left (right) translations defining finite gauge transformations\footnote{A bit later the authors of \cite{DiCosmo:2023wth} obtained finite Poisson gauge transformations integrating the sections of the corresponding Lie algebroid.}. Using the reach geometry of symplectic groupoids, it is also possible to define gauge-covariant and gauge-invariant strength tensors of the electromagnetic field as well as to introduce a minimal electromagnetic coupling to point particles and  matter fields \cite{sharapov2024poisson}. All in all, the concept of symplectic groupoids looks very promising for studying  noncommutative dynamics of fields and particles in semi-classical approximation.

It should be noted that the construction of a symplectic groupoid that integrates a given Poisson structure is a difficult mathematical problem, whose explicit solution is known only in some special cases.  The first nontrivial and best-studied case is that of linear Poisson brackets mentioned above. Therefore, in this paper, we focus on  the classical dynamics of point particles whose position `coordinates' form a Lie algebra. In semi-classical approximation, this gives rise to Hamiltonian dynamics on the cotangent bundle $T^\ast G$ of the corresponding Lie group $G$.  A typical example of such a Hamiltonian system is the motion of a rigid body about a fixed point. Here, the configuration space is given by the rotation group $SO(3)$, whose cotangent bundle -- the phase space of the system -- is diffeomorphic to $SO(3)\times \mathbb{R}^3$. This means that, formally, a nonrelativistic particle moving in $3$-space with 
$\mathfrak{so}(3)$ noncommutativity and a rigid body with a stationary point inside the body have the same phase space. The fundamental difference, however, lies in the physical interpretation of the phase space variables: what is the configuration space for the rigid body becomes the momentum space for the particle and vice versa. 
The compactness of the rotation group $SO(3)$, considered as momentum space\footnote{In Sec. \ref{S4},  we replace $SO(3)$ with its universal covering $SU(2)$, which is also compact.}, has far-reaching consequences.  For example, the $3$-velocity of a free (nonrelativistic!) particle turns out to be bounded from above, provided the kinetic energy is a smooth function of momenta. In Sec. \ref{secSG}, we consider the motion of such a particle in a central potential and show that
it cannot fall to the centre, whatever attractive force. For any repulsive central force, we find a domain of bounded motion. All these observations contradict the usual physical intuition. Sec. \ref{S4} is devoted to a detailed study of Kepler's problem for a particle with $\mathfrak{so}(3)\simeq \mathfrak{su}(2)$ noncommutativity of space coordinates. As momentum space, we take the group manifold $SU(2)\sim \mathbb{S}^3$, so that the total phase space is given by the product $\mathbb{R}^3\times \mathbb{S}^3$. We show that all space trajectories are plane curves provided the kinetic energy is invariant under rotations. However, unlike the commutative case, the planes do not pass through the centre of force. Finally, we present a Hamiltonian for which all spatial trajectories are conical curves, as in the commutative Kepler problem.

\section{Noncommutative mechanics in semi-classical approximation}
Noncommutative geometry studies noncommutative algebras as if they were algebras of functions on spaces, like, e.g., the commutative algebras  associated with smooth manifolds. Such geometric models for noncommutative algebras provide a  natural mathematical framework for classical and quantum field theory in 
noncommutative spacetime. The correspondence principle between commutative and noncommutative physics implies that appropriate noncommutative algebras must be close to the usual commutative algebras of smooth functions, at least starting from some fundamental scale $\ell$.  
This, in turn, suggests the existence of a one-parameter family of algebras $\mathfrak{A}_\ell$ such that $\mathfrak{A}_0$ is isomorphic to the algebra of smooth functions on a given spacetime manifold $X$. In other words, $\mathfrak{A}_\ell$ can be considered as a deformation of the commutative algebra $C^\infty(X)$, with the deformation parameter 
$\ell\geq 0$ controlling  the degree of non-commutativity. The product in $\mathfrak{A}_\ell$, denoted by $\star$, should then have the form
\begin{eqnarray}\label{star}
    f\star g=f\cdot g+\frac{\ell}{2}\{f,g\}+\mathcal{O}(\ell^2)\qquad \forall f,g\in C^{\infty}(X)\,.
\end{eqnarray}
If we further assume that the $\star$-product algebra is associative for all $\ell$, then the bilinear operator determining the first-order deformation must be equivalent to some Poisson bracket on $X$.  In terms of local coordinates $x^a$, each Poisson bracket on $X$ defines and is defined by a Poisson bivector
\begin{equation}\label{PB}
    \Theta^{ab}(x)=\{x^a, x^b\}\,.
\end{equation}
The celebrated Kontsevich formality theorem \cite{Kontsevich:1997vb} ensures the existence of higher-order terms in (\ref{star}), complementing the first-order deformation to an associative product. Using the terminology of quantum mechanics, we will refer to the Poisson manifold $(X,\Theta)$ as a semi-classical approximation to noncommutative spacetime (= the algebra $\mathfrak{A}_\ell$).  It is this approximation that we will deal with in this paper.

Regarding now the Poisson manifold $X$ as a configuration space of a point scalar particle, we can ask about the corresponding phase space, that is, the space of states of the particle.  With the correspondence principle in mind, we should expect this to be given by a symplectic manifold $\mathcal{G}$ whose dimension is equal to twice the dimension of the configuration space $X$. The compatibility between phase space and configuration space dynamics implies that the Poisson brackets on $\mathcal{G}$, associated with the symplectic structure, should reproduce the initial Poisson brackets (\ref{PB}) on functions that depend exclusively on the spacetime coordinates $x^a$. Among other things, this means that the symplectic manifold $\mathcal{G}$ is a fibre bundle over the Poisson manifold $X$ whose projection $\sfs: \mathcal{G}\rightarrow X$ is a Poisson map.  This is known as a symplectic realization of a Poisson manifold $(X,\Theta)$. 

In general, finding a symplectic realization for a given symplectic manifold is a difficult problem. Moreover, the problem may have different solutions in different dimensions \cite{Weinstein,Gracia-Bondia:2001ynb}, which makes the definition of a particle's phase space quite ambiguous.  In \cite{kupriyanov2024symplectic}, it was suggested to narrow the ambiguity by requiring the phase space to be a {\it symplectic groupoid} integrating the Poisson bracket on $X$. Here we will refrain from discussing symplectic groupoids in any detail, referring the interested reader to the books \cite{KM, vaisman2012lectures, CFM}. In a nutshell, a symplectic groupoid $\mathcal{G}$ over a Poisson manifold $X$ is a symplectic manifold of $\dim \mathcal{G}=2\dim X$ endowed with a compatible groupoid structure. As part of this structure, there exists a pair of canonical projections $\sfs, \sft: \mathcal G\rightarrow X$, called the {\it source} and {\it target},  each of which makes $\mathcal{G}$ into a fibre bundle over $X$. It is also required that the source map $\sfs$ be Poisson and the target map $\sft$ be anti-Poisson: 
\begin{equation}
    \{\sfs^\ast f,\sfs^\ast g\}_{\mathcal{G}}=\{f,g\}_X\,,\qquad \{\sft^\ast f, \sft^\ast g\}_{\mathcal{G}}=-\{f,g\}_X\qquad \forall f,g\in C^{\infty}(X)\,.
\end{equation}
The existence of two canonical projections is stressed in the shorthand notation $\mathcal{G}\rightrightarrows X$ for a groupoid $\mathcal G$ over $X$. One more ingredient of the groupoid structure is an involutive map  $\sfi: \mathcal{G}\rightarrow \mathcal{G}$, called {\it inversion}, that interchanges source and target projection: $\sfs\circ \sfi=\sft$ and  $\sft\circ \sfi=\sfs$. From the viewpoint of classical mechanics, the fibre  $\sfs^{-1}(x)$ is  the space of (generalized) momenta at $x\in X$.

A trivial example of symplectic groupoids is provided by the cotangent bundle $T^\ast X$ endowed with the canonical symplectic structure. This groupoid
integrates the zero Poisson bracket on $X$ and describes the phase space of an ordinary mechanical system. Here the source and target maps coincide with 
each other and with the canonical projection ${p}: T^\ast X\rightarrow X$.  The inversion acts by inverting the direction of canonical momenta.

Notice that whatever configuration space $X$, the generalised momenta of an ordinary mechanical system always form a linear space, the cotangent space $T^\ast_xX$ at $x\in X$. For a general symplectic groupoid $\mathcal{G}\rightrightarrows X$ the role of a momentum space at $x\in X$ is played by
the fibre $\sfs^{-1}(x)\subset \mathcal{G}$, which may well be a manifold with nontrivial topology. The possibility of a curved momentum space sharply distinguishes noncommutative classical mechanics from its commutative counterpart and has far-reaching dynamical implications as we will see soon.

In general, not every Poisson manifold $(X,\Theta)$ can be integrated into a smooth symplectic groupoid $\mathcal{G}\rightrightarrows X$. 
For such integration to be possible, the Poisson manifold should meet some integrability conditions \cite{CFM}. When the conditions are satisfied, the corresponding Poisson structure is called {\it integrable}. Significantly, every integrable Poisson structure gives rise to a symplectic groupoid, which is unique up to covering.

Given a symplectic groupoid $\mathcal{G}\rightrightarrows X$, one defines dynamics by specifying a Hamiltonian $H\in C^\infty(\mathcal{G})$. The time evolution of any classical observable $f\in C^\infty(\mathcal{G})$ is determined by the Hamiltonian equation of motion
\begin{equation}
    \dot f=\{f,H\}_{\mathcal{G}}\,.
\end{equation}
Equivalently, one can describe the phase-space trajectories $\gamma:\mathbb{R}\rightarrow \mathcal{G}$ as stationary points of the action functional
\begin{equation}\label{act}
    S=-\int_\gamma \left[\theta +H\dd t\right]\,,
\end{equation}
where $\theta$ is a symplectic potential for the symplectic form $\omega=\dd \theta$ on $\mathcal{G}$.

\section{Symplectic groupoids and dynamics for Lie--Poisson structures \label{secSG}}

Let us specify the above considerations for the case of linear Poisson bracket
\begin{equation}\label{LP}
    \{x^a,x^b\}=f^{ab}_cx^c\,.
\end{equation}
Here, $x^a$ are Cartesian coordinates on $X=\mathbb{R}^n$. It follows immediately from the Jacobi identity that $f_c^{ab}$ are structure constants of some $n$-dimensional  Lie algebra $\mathfrak{g}$.  This allows us to identify $\mathbb{R}^n$ with the space dual to $\mathfrak{g}$. The linear Poisson bracket (\ref{LP}) on $\mathfrak{g}^\ast$ is known as the Lie--Poisson structure. 

To any Lie algebra $\mathfrak{g}$ there corresponds a Lie group $G$, which is unique up to covering.  As we will see in a moment, the group $G$  plays the role of the space of momenta conjugate to the position coordinates $x^a$. To stress this interpretation, we will denote local coordinates on $G$ by $p_i$. We define the basis $\bar{\gamma}^a$ of left-invariant $1$-forms on $G$ by the standard relations
\begin{equation}
    g^{-1}\dd g=\bar \gamma_a t^a\,, \qquad \bar{\gamma}_a=\bar{\gamma}_a^i(p)\dd p_i\,,
\end{equation}
where $g=g(p)\in G$ and $t^a$ is a basis in $\mathfrak{g}$ with commutation relations $[t^a,t^b]=f^{ab}_ct^c$. The dual to the basis of left-invariant $1$-forms $\{\bar\gamma_a\}$ is the basis $\{\gamma^a\}$ of left-invariant vector fields on $G$. These satisfy the relations 
\begin{equation}
\bar\gamma_a(\gamma^b)=\delta^b_a\,,\qquad   \dd \bar\gamma_a=-\frac12 f^{bc}_a\bar\gamma_b\wedge \bar\gamma_c\,,\qquad [\gamma^a,\gamma^b]=f^{ab}_c\gamma^c\,.
\end{equation}
Now we are ready to define a symplectic groupoid that integrates the linear Poisson bracket (\ref{LP}). As a smooth manifold, it is given by the Cartesian product $\mathcal{G}=\mathfrak{g}^\ast\times G$ and the symplectic structure is determined by the exact $2$-form 
\begin{equation}\label{LF}
    \omega =\dd \theta \,,\qquad \theta =\langle x, g^{-1}\dd g\rangle =x^a \bar\gamma_a^i(p)\dd p_i
\end{equation}
or, more explicitly,
\begin{equation}
    \omega=\dd x^a\wedge \bar\gamma_a-\frac12x^cf^{ab}_c\bar\gamma_a\wedge\bar\gamma_b\,.
\end{equation}
The bivector dual to $\omega$ defines the Poisson brackets 
\begin{equation}
    \{x^a,x^b\}=f^{ab}_cx^c\,,\qquad \{x^a,p_i\}=\gamma^a_i(p)\,,\qquad \{p_i,p_j\}=0\,. 
\end{equation}
The source, target, and inversion maps are given by
\begin{equation}
    \sfs(x,g)=x\,,\qquad \sft(x,g)=\mathrm{Ad}^\ast_g(x)\,, \qquad \sfi(x,g)=(\mathrm{Ad}^\ast_g(x),g^{-1})\,,
\end{equation}
where $(x,g)\in \mathfrak{g}^\ast\times G$ and $\mathrm{Ad}^\ast$ stands for the coadjoint representation of the group $G$. 

The action functional (\ref{act}) for a nonrelativistic particle in noncommutative space $\mathfrak{g}^\ast$ assumes now the form 
\begin{equation}\label{S}
    S[x, g]=-\int \big[\langle x,  g^{-1}\dot g \rangle + H(x,g)\big]dt\,.
\end{equation}
The least action principle gives the Hamiltonian equations of motion 
\begin{equation}\label{HE}
    \dot x^a=\{x^a, H\}=\gamma^a_i(p) \frac{\partial H}{\partial p_i}+x^c\,f^{ab}_c\frac{\partial H}{\partial x^b}\,,\qquad \dot p_i=\{p_i, H\}=-\gamma_i^a(p)\frac{\partial H}{\partial x^a}\,.
\end{equation}

The direct product structure of the phase space $\mathfrak{g}^\ast\times G$ allows one to consider Hamiltonians that are given by a sum of kinetic and potential energy,   $H=T(g)+U(x)$.  Such mechanical systems are usually called {\it natural}. For a free particle, $U(x)=0$ and Eqs. (\ref{HE}) assume very simple form 
\begin{equation}\label{fEq}
    \dot x^a=\gamma^a_i(p) \frac{\partial T}{\partial p_i}\,,\qquad \dot p_i=0\,.
\end{equation}
The first equation expresses the particle's velocity $\dot x^a$ as a function of its momentum $p_i$. The second equation says that the momentum of a free particle is conserved. By the Noether theorem, the Hamiltonian action of the integrals of motion $p_i$ generates the global symmetry of the free action:
\begin{equation}\label{tr}
    x^a\rightarrow x^a+ v^i \gamma^a_i(p)\,,\qquad \forall v^i=const\,.
\end{equation}
The general solution to (\ref{fEq}) reads 
\begin{equation}
    x^a(t)=x^a(0)+\dot x^a(0)t\,.
\end{equation}
Whatever the group $G$, we get a uniform linear motion in space, as for ordinary free particles. 
Therefore, it seems impossible to distinguish between commutative and noncommutative spaces by probing them with free particles. 
The only exception is the case of compact $G$: because of the first equation in (\ref{fEq}) the components of the velocity vector $\dot x^a $ appear to  be bounded by absolute value  (like in Special Relativity)
provided  the kinetic energy $T(p)$ has continuous first derivatives.

Now consider a particle moving in response to an external force. In this case, the equations of motion take the form 
\begin{equation}
    \dot x^a=\gamma^a_i(p) \frac{\partial T}{\partial p_i}+x^c\,f^{ab}_c\frac{\partial U}{\partial x^b}\,,\qquad \dot p_i=-\gamma_i^a(p)\frac{\partial U}{\partial x^a}\,.
\end{equation}
(Notice that due to the external force, the velocity $\dot x$ may now be unbounded even for a compact $G$.) 
The co-adjoint action of $G$ on $\mathfrak{g}^\ast$ stratifies $\mathfrak{g}^\ast$ into co-adjoint orbits. Similarly, the action of $G$ on itself by conjugation stratifies $G$ into the classes of conjugate elements. Let us suppose that the functions $T$ and $U$ are invariant under these actions, that is, $U$ is constant on each co-adjoint orbit, while $T$ is constant on each conjugacy class:
\begin{equation}\label{UTinv}
    U(x)=U(\mathrm{Ad}^\ast_h(x))\,,\qquad T(g)=T(hgh^{-1})\,,\qquad \forall h\in G\,.
\end{equation}
In mathematics, such functions $T$ are called {\it class functions}. Then one can readily see that the action 
\begin{equation}\label{STU}
    S[x, g]=-\int \big(\langle x, g^{-1}\dot g\rangle + T(g)+U(x)\big)dt
\end{equation}
is invariant under the transformations
\begin{equation}
    x\rightarrow \mathrm{Ad}^\ast_h x\,,\qquad g\rightarrow hgh^{-1}\,,\qquad \forall h\in G\,.
\end{equation}
Again, by the Noether theorem, these symmetry transformations give rise to conservation laws. To write the corresponding conserved quantities in an explicit form we need to introduce the right-invariant $1$-forms and  vector fields on $G$:
\begin{equation}
    dgg^{-1}=t^a \rho^i_a(p)dp_i\,,\qquad \bar \rho^a=\bar \rho^a_i(p)\frac{\partial}{\partial p_i}\,,\qquad  \bar \rho^a(\rho_b)=\delta_b^a\,.
\end{equation}
They satisfy the identities 
\begin{equation}
    d\rho_c=\frac12f_c^{ab}\rho_a\wedge \rho_b\,,\qquad 
  [\bar \rho^a,\bar \rho^b]=-f^{ab}_c\bar\rho^c\,,\qquad [\bar \rho^a, \gamma^b]=0\,. 
  \end{equation}
We also introduce the matrix $\Delta^a_b(p)=\bar\gamma_b(\bar \rho^a)$, which obeys the equation
\begin{equation}
\gamma^c\Delta^a_b=-f^{cd}_b\Delta_d^a\,.
\end{equation}
Conditions (\ref{UTinv}) are equivalent to
\begin{equation}
x^bf_b^{ac} \frac{\partial U}{\partial x^c}=0\,,\qquad (\gamma^a-\bar \rho^a)T=0\,.
\end{equation}
We claim that, under the conditions above,  the following functions are conserved:
\begin{equation}\label{mu1}
\mu^a=x^b \bar\gamma_b^i(\gamma^a_i-\bar \rho_i^a)=x^a-x^b\Delta_b^a(p)\,.
\end{equation}
Indeed, 
\begin{equation}
\begin{array}{c}
    \dot \mu^a=\dot x^b \bar\gamma_b^i(\gamma^a_i-\bar \rho_i^a)-\dot p_i\partial^i_p\Delta_b^a(p)  x^b \\[5mm]
  \displaystyle  =\frac{\partial T}{\partial p_i}(\gamma^a_i-\bar \rho_i^a)+\frac{\partial U}{\partial x^c}
  (\gamma^c\Delta^a_b) x^b=\frac{\partial T}{\partial p_i}(\gamma^a_i-\bar \rho_i^a)-\frac{\partial U}{\partial x^c}f^{cd}_bx^b\Delta_d^a =0\,.   
\end{array}
\end{equation}
The conserved quantities $\mu^a$ have the following Poisson brackets:
\begin{equation}
    \{\mu^a,\mu^b\}=f^{ab}_c\mu^c\,.
\end{equation}
If $G$ is a semi-simple Lie group, then the Killing metric $g^{ab}=f^{ac}_df^{db}_c$ is nondegenerate and we can define the conserved quantity $Q=g_{ab}\mu^a\mu^b$, which satisfies $ \{Q, \mu^a\}=0$.

As mentioned above, for a compact $G$, the velocity of a free non-relativistic particle turns out to be bounded above. However, this property may not survive upon switching on interaction.  We conclude this section by discussing two more interesting phenomena related to the compactness of the momentum space $G$. 
The phase-space dynamics develop on the isoenergetic surface 
\begin{equation}\label{TUE}
    T(g)+U(x)=E
\end{equation}
defined by some energy level $E$. For a compact momentum space, the kinetic energy necessarily reaches its maximum and minimum values:
\begin{equation}
    T_{min}\leq T(g)\leq T_{max}\,.
\end{equation}
It is the existence of an upper bound on the kinetic energy that sharply differs the situation under consideration from  ordinary classical mechanics. On substituting (\ref{TUE}), we get
\begin{equation}
   E-T_{max} \leq U(x)\leq E-T_{min}\,.
\end{equation}
 Depending on $U$, these inequalities can severely bound the domain of motion.  Suppose, for example, that the potential energy of a particle drops to zero at infinity, i.e.,
\begin{equation}
    U(x)\rightarrow 0\,,\qquad x\rightarrow \infty\,.
\end{equation}
Then the particle has to move in a {\it compact} region of space
\begin{equation}\label{CE}
   \mathcal{C}_E=\big\{ x\in \mathfrak{g}^\ast\;\big|\;U(x)\geq E-T_{max} \big\}
\end{equation}
whenever its total energy $E>T_{max}$. The last inequality is automatically fulfilled for all points of the bounded region\footnote{$\mathcal{C}$ stands for {\it compact} or {\it confinement}. } 
\begin{equation}\label{Cdef}
   \mathcal{ C}=\big\{ x\in \mathfrak{g^\ast}\;\big|\;U(x) > T_{max}-T_{min} \big\}\,.
    \end{equation}
    (The case $\mathcal{C}=\varnothing$ is not excluded.) 
Therefore, if a  trajectory starts at some point $x\in \mathcal{C}$, it will be fully confined in a bounded region of $ \mathfrak{g}^\ast$. Remarkably, the confinement may take place even for {\it repulsive} central force  vanishing at infinity! In Sec. \ref{ST}, we will illustrate this phenomenon with the Coulomb potential. 

Similarly, we can consider an {\it attractive} central force that becomes infinite at some point $a$, i.e., 
\begin{equation}
    U(x)\rightarrow -\infty\,, \qquad x\rightarrow a \,. 
\end{equation}
Normally, we would expect some trajectories to lead to the singular point $x=a$ (falling to the centre). As the particle approaches the centre, the decrease in potential energy should be balanced by the increase in kinetic energy.
The domain available for motion is still described by Eq. (\ref{CE}). Now it means that the particle cannot approach the singular point $a$ arbitrarily close; whatever total energy $E$, the region where 
\begin{equation}\label{nofallregion}
    U(x)\leq E-T_{max}
\end{equation}
is forbidden for motion. This inequality does not exclude the physical situation where an accelerating charge loses its energy due to electromagnetic radiation and spirals towards the centre. In this paper, however, we do not consider the radiation effects. 

{ In conclusion, we note that the abnormal behavior above owes its existence entirely to the compactness of momentum space and can occur even in a commutative situation. As shown in \cite{A4}, for one, the toric compactification of the momentum space associated with the commutative configuration space $\mathbb{R}^n$ eliminates the distinction between attractive and repulsive potentials.}

\section{Example: a scalar particle in  $\mathfrak{su}(2)$ noncommutative space}\label{S4}

Let us illustrate the above formulas by a nonrelativistic particle moving in the noncommutative $3$-space $\mathfrak{su}(2)^\ast$.  {To our knowledge, this mechanical model was first studied in \cite{A1,A2,A3}}.   Note that $\mathfrak{su}(2)$  is a unique compact simple Lie algebra of dimension three. As the space of momenta of the particle, we  take the simply-connected Lie group $SU(2)$, so that the whole phase space is given by the direct product $\mathcal{G}=\mathfrak{su}(2)^\ast\times SU(2)$. To write the Hamiltonian equations of motion on $\mathcal{G}$ it is convenient to use the language of quaternions. 

Recall that the quaternion algebra $\hh$ is a unique $4$-dimensional associative  division algebra over the reals. As a vector space $\hh$ is spanned by the multiplicative identity element $1$ together with the three `imaginary units' $\mathbf{i}$, $\mathbf{j}$, and $\mathbf{k}$ subject to the relations
\begin{align*}
    \mathbf{i}^2=\mathbf{j}^2=\mathbf{k}^2=\mathbf{i}\,\mathbf{j}\,\mathbf{k}=-1 \ .
\end{align*}
Hence, the general element of $\hh$ is
\begin{equation}
    q=q_0+q_1\mathbf{i}+q_2\mathbf{j}+q_3\mathbf{k}\,, \qquad q_0,q_1,q_2,q_3\in \mathbb{R}\,.
\end{equation}
The conjugation, norm, and trace on $\hh$ are defined  by the usual formulas:
\begin{equation}\label{CNT}
\begin{array}{l}
    \bar q=q_0-q_1\mathbf{i}-q_2\mathbf{j}-q_3\mathbf{k}\,,\\[3mm]
    |q|^2=q\bar q=q_0^2+q_1^2+q_2^2+q_3^2\,,\\[3mm] 
    \mathrm{Tr}(q)=q+\bar q=2q_0\,.
    \end{array}
\end{equation}
These have the standard properties 
\begin{equation}
  \overline{qq'}=\bar q'\bar q\,,\qquad  |q q'|=|q||q'|\,,\qquad \mathrm{Tr}(qq')=\mathrm{Tr}(q' q)\,.
\end{equation}

In terms of quaternions, the Lie algebra $\mathfrak{su}(2)$ is identified with the commutator algebra of purely imaginary quaternions, while the Lie group $SU(2)$ can be described as the multiplicative group of quaternions of norm $1$. Using the Eucledian structure on the space of imaginary quaternions, we can also identify $\mathfrak{su}(2)$ and $\mathfrak{su}(2)^\ast$.  Then the coadjoint action of $SU(2)$ on the physical $3$-space $\mathfrak{su}(2)^\ast$ is given simply by conjugation
\begin{equation}
    \mathsf{Ad}^\ast_u(\mathbf {x})=u\mathbf{x}\bar u,
\end{equation}
where $\mathbf{x}+\bar {\mathbf{x}}=0$ and $u\bar u=1$. In the following, we will denote pure imaginary quaternions in bold. 

The presymplectic potential (\ref{LF}) on $\mathfrak{su}(2)^\ast\times SU(2)$ takes  now the form 
\begin{equation}
    \theta =\mathrm{Tr}(x\bar u\dd u)
\end{equation}
and for the symplectic $2$-form we get
\begin{equation}
    \omega=\dd \theta =\mathrm{Tr}\big(\dd x\wedge \bar u \dd u- x\bar u\dd u\wedge \bar u\dd u\big)\,.
    \end{equation}
Instead of working with constrained momenta $u$, we can add the  constraint $u\bar u-1=0$ to the Hamiltonian $ H(u,x)$ with the Lagrange multiplier $\lambda$. Then the full action (\ref{S}) with unconstrained $u$ is given by  
\begin{equation}\label{SUU}
    S[\mathbf{x},u, \lambda]=\int \Big(-\frac12\mathrm{Tr}(\mathbf{ x}\bar u\dot u)-H(\mathbf{x}, u)-\lambda(u\bar u-1)\Big)\dd t\,.
\end{equation}
Let us take the Hamiltonian in the form  ${H}=T(u)+U(\mathbf{x})$  and introduce the pair of quaternions
\begin{equation}
   V_0+ \mathbf{V}=\frac{\partial T}{\partial u_0}+\frac{\partial T}{\partial u_1}\mathbf{i}+\frac{\partial T}{\partial u_2}\mathbf{j} +
    \frac{\partial T}{\partial u_3}\mathbf{k}\,,\qquad  \mathbf{F}=-\frac{\partial U}{\partial x_1}\mathbf{i}-\frac{\partial U}{\partial x_2}\mathbf{j} -
    \frac{\partial U}{\partial x_3}\mathbf{k}\,.
    \end{equation}
The variation of  (\ref{SUU}) results in the  equations of motion
\begin{equation}\label{EoM}
\dot {{u}}= u\mathbf{F}\,,\qquad \dot{\mathbf{x}}=u_0\mathbf{V}-V_0\mathbf{u}+\frac 12[\mathbf{V},\mathbf{u}]+[\mathbf{x}, \mathbf{F}]\,,
\qquad |u|^2=1\,.
\end{equation}
Here, the square brackets stand for the commutator of quaternions. Notice that the algebraic constraint $|u|^2=1$ is conserved over time.

The equations of motion (\ref{EoM}) are considerably simplified if we assume that $U$ is a central potential of the form $U=f(|\mathbf{x}|^2)$ and take $T=-u_0/\ell m$. In this case,
\begin{equation}
    \dot u_0=-\mathrm{Tr}(\mathbf{u}\mathbf{x})f'\,,\qquad   \dot {\mathbf{u}}=\big([\mathbf{x}, \mathbf{u}]-2u_0\mathbf{x} \big)f'  \,,\qquad \ell m\dot{\mathbf{x}}=\mathbf{u}\,, \qquad  |u|^2=1\,.
\end{equation}
We see that the direction of the radial acceleration depends on the sign of $u_0$, regardless of whether the central potential is attractive or repulsive; there is also an additional  force proportional to the angular momentum of the particle.  Excluding $\mathbf{u}$, we obtain the system of first- and second-order equations 
\begin{equation}\label{ddx}
    \ddot{\mathbf{x}}=(2T\mathbf{x} +[\mathbf{x},\dot{\mathbf{x}}])f'\,, \qquad \dot T=-\mathrm{Tr}(\dot{\mathbf{x}}\mathbf{x})f'
    \end{equation}
with initial data lying on the constraint surface $T^2+|\dot{\mathbf{x}}|^2=(\ell m)^{-2}$. Since the constraint is conserved over time, the absolute values of the kinetic energy and velocity appear to be bounded above, $|T|, |\dot{\mathbf{x}}|\leq (\ell m)^{-1} $. For small velocities, when $|\dot{\mathbf{x}}|<<(\ell m)^{-1}$ and $T\approx -(\ell m)^{-1}$, the first equation in (\ref{ddx}) turns to the standard Newton equation for a particle of mass $m$ moving in the external potential $U$. As for the second equation, it represents the law of energy conservation $T+U=E$, which allows us to rewrite the system (\ref{ddx}) in the form 
\begin{equation}\label{ddxE}
    \ddot{\mathbf{x}}=\big(2(E-f)\mathbf{x} +[\mathbf{x},\dot{\mathbf{x}}]\big)f'\,,
    \end{equation}
 where the constant $E$ is determined by the initial data through the equation
 $$\big (E-f(|\mathbf{x}|^2)\big)^2+|\dot{\mathbf{x}}|^2=(\ell m)^{-2}\,.$$
One more integral of motion is given by the `angular momentum' (\ref{mu1}). In terms of quaternions it can be written as
\begin{equation}\label{mu}
    {\mu}=\mathbf{x}-u\mathbf{x}\bar u\,.
\end{equation}

\begin{remark}
 We see that a free particle with $SU(2)$ as momentum space and a class function $T(u_0)$ as Hamiltonian has the same number of fundamental (time-independent) conservation lows as the ordinary nonrelativistic particle. These are given by the three independent components of the momentum $u$, the angular momentum vector $\mu$, and the kinetic energy $T(u_0)$. All these quantities generate and come from symmetry transformations. Furthermore, free motion is uniform and linear. 
What about the remaining symmetry transformations of Newton's mechanics, namely,  the Galilean boosts: $x\rightarrow x+tv$?  One can see that under mild assumptions, they are implicitly present in the free model (\ref{STU}) for any Lie group $G$. Let us suppose that the kinetic energy $T(g)$ has a nondegenerate local minimum at the identity element  $e\in G$, i.e., $\partial_iT (e)=0 $ and the matrix $\big(\partial_i\partial_j T(e)\big)$ is positive-definite. By the Morse lemma, one can find new coordinates $P^i=P^i(g)$ in a vicinity of $e$ such that $T=T(e)+\frac12\delta_{ij}P^iP^j$. In these coordinates, the free Lagrangian (\ref{STU})  takes the form
$$
L=-x_a\bar{\gamma}^a_i(P)\dot P^i-\frac12\delta_{ij}P^iP^j\,.
$$
(The constant $T(e)$ can be omitted.) Introducing the new position coordinates $X_i=x_a\bar{\gamma}_i^a(P)$, we can bring the Lagrangian into the standard form:
$$
L=\dot X_i P^i-\frac12\delta_{ij}P^iP^j\,.
$$
The last Lagrangian is invariant (up to a total derivative) under the Galilean boosts: $X\rightarrow X+ Vt$ and $P\rightarrow P+V$.
By making the inverse change of variables $(X,P)\rightarrow (x,g)$, we get  nonlinear symmetry transformations for the free action (\ref{STU}). One can regard $(X, P)$ as Darboux coordinates on $T^\ast G$ adapted to the Hamiltonian $T$. 
\end{remark}

To further elucidate the connection with ordinary Hamiltonian mechanics, it is convenient to introduce dimensional momenta.   If $p$ is the momentum canonically conjugated to a position coordinate $x$, then 
$[x][p]=[\hbar]$. 
In the unit system where $\hbar=1$, we have $[p]=[x]^{-1}$. Let $\ell$ be a characteristic length scale,  $[\ell]=[x]$. Introduce the dimensional momenta $p$ as $u=\ell p\in \mathbb{H}$.  For the action functional $S$ to have the dimension of $\hbar$ we also need to multiply the first term in (\ref{SUU}) by $\ell^{-1}$. The Hamiltonian for a nonrelativistic particle of mass $m$ can be written as
\begin{equation}
    \mathcal{H}=\frac{k(u_0)}{m\ell^2}+U(\mathbf{x})\,,
\end{equation}
where $k(u_0)$ is a dimensionless quantity. 
Due to the constraint $|p|^2=\ell^{-2}$, we can parameterize the southern hemisphere of $\mathbb{S}^3$ by the vector part of the quaternion $p=p_0+{\mathbf{ p}}$. 
In these coordinates, the first-order Lagrangian assumes the form 
\begin{equation}\label{Lxp}
\begin{array}{c}
\displaystyle L= -\frac{\ell}{2}\mathrm{Tr}(\mathbf{x} \bar p \dot  p)-\mathcal{H} \\[5mm]
\displaystyle =\frac12\mathrm{Tr}({\mathbf{p}}\dot{\mathbf{ x}})\sqrt{1-\ell^2|\mathbf{ p}|^2} +\frac{\ell}{2} \mathrm{Tr}\big(\mathbf{x} \mathbf{ p} \dot{\mathbf{p}}\big) +\frac{\ell^2}{2}\frac{\mathrm{Tr}(\mathbf{x}\mathbf{p})\mathrm{Tr}({\mathbf{ p}}\dot{\mathbf p})}{\sqrt{1-\ell^2|\mathbf{p}|^2}}\\[5mm]
\displaystyle-\frac{k(\sqrt{1-\ell^2|\mathbf{ p}|^2})}{m\ell^{2}}-U(\mathbf{x})+\text{tot. der.}
\end{array}
\end{equation}
The terms with time derivatives determine the symplectic potential, which leads to the following Poisson brackets:
\begin{equation}
    \{x^i,x^j\}=2\,\ell\,\varepsilon^{ijk}x^k\,,\qquad \{x^i,p_j\}=\sqrt{1-\ell^2|\mathbf{p}|^2}\,\delta^i_j+\ell \,\varepsilon^{ijk}p_k\,,\qquad \{p_i,p_j\}=0\,.
\end{equation}
We see that the commutative limit $\ell\rightarrow 0$ for the space coordinates is simultaneously the decompactification limit  for the momentum $3$-sphere $\mathbb{S}^3$. In  this limit, the compact nonabelian group $SU(2)$ turns into the additive group of the vector space $\mathbb{R}^3$. Expanding the Lagrangian (\ref{Lxp}) in powers of $\ell$, we find

\begin{equation}\label{LLxp}
    L= \frac12\mathrm{Tr}(\mathbf{p}\dot {\mathbf{x}})+\frac{\ell}{2} \mathrm{Tr}({\mathbf  x} \mathbf{ p}\dot{\mathbf{p}})
-\frac{k(1)}{m\ell^2}+k'(1)\frac{|\mathbf{p}|^2}{2m}  - U(\mathbf{x})+\mathcal{O}(\ell^2)\,.
\end{equation}
We can drop the constant term as it does not affect dynamics. Then (\ref{LLxp}) goes into the standard Lagrangian  of a nonrelativistic particle as $\ell\rightarrow 0$ provided that $k'(1)=-1$. For example, we can set as above $k=-u_0$. For $\ell\neq 0$, the corrections due to noncommutativity become negligible whenever the energy-momentum $p$ of the particle lies in a small vicinity of the identity element of $SU(2)$, that is, $ p\approx \ell^{-1}$. One can use the Lagrangian (\ref{LLxp}) to derive the leading corrections to ordinary nonrelativistic dynamics due to space noncommutativity. We find 
\begin{equation}
m\dot{\mathbf{x}}=\mathbf{p}-m\ell [\mathbf{x},\dot{\mathbf{p}}]   +\mathcal{O}(\ell^2)\,,\qquad \dot{\mathbf{p}}=\mathbf{F}
+\frac{\ell}2[\mathbf{p},\dot{\mathbf{p}}]+\mathcal{O}(\ell^2)
\end{equation}
or
\begin{equation}
 m\ddot{\mathbf{x}}=\mathbf{F}+\frac{ m\ell}{2} [\dot{\mathbf{x}}, \mathbf{F}]-m\ell \frac{d}{dt}[\mathbf{x},\mathbf{F} ]+\mathcal{O}(\ell^2)\,.
\end{equation}
The last term vanishes for central potentials. For instance, if $U=\alpha/|\mathbf{x}|$ is the Coulomb potential, then 
the equation becomes
\begin{equation}\label{ddx2}
 m\ddot{\mathbf{x}}=\alpha\frac{\mathbf{x}}{\;|\mathbf{x}|^3}+\frac{ \alpha m\ell}{2} \frac{[\dot{\mathbf{x}}, \mathbf{x}]}{\;|\mathbf{x}|^3}+\mathcal{O}(\ell^2)\,.
\end{equation}
Notice the term proportional to the angular momentum of the particle. This looks like a  force exerted on the charge by a magnetic monopole located at the origin. A nontrivial point is that the magnitude of this additional force is proportional to the {\it mass} of the charged particle. {  As with the case of gravity, this indicates that the force is of geometric origin}. Expression (\ref{mu}) gives the following integral of motion:
\begin{equation}\label{mumu}
    \mu= m[\mathbf{x},\dot{\mathbf{x}}]-2\alpha m\ell \frac{\mathbf{x}}{|\mathbf{x}|}+\mathcal{O}(\ell^2)\,.
\end{equation}
It is noteworthy that the vector $\mu$ is exactly conserved if one restricts to the leading corrections in (\ref{ddx2}) and (\ref{mumu}). In this approximation, $\mu$ coinsides with the Poincar\'e vector, which allows one to find exact trajectories of an electric charge moving in the dyon potential, see e.g.  \cite{2000EJPh21183S}.

\section{Noncommutative Kepler problem in semi-classical limit}

In this section, we consider in some detail the motion of a point charge $q$ in the Coulomb potential centred at the origin of the $\mathfrak{su}(2)$ noncommutative space. It may be shown \cite{Kurkov:2021kxa,Kupriyanov:2023gjj} that  the standard Coulomb potential, i.e., the field configuration $\vec A= 0$ and $A_0=C/|\mathbf{x}|$, solves the field equations of Poisson electrodynamics with $\mathfrak{su}(2)$ non-commutativity in the whole space except the origin.  Applying the recipe of minimal interaction proposed in Ref. \cite{kupriyanov2024symplectic} allows one to couple a charged particle on  $\mathfrak{su}(2)$ noncommutative space to the Coulomb field. In the electrostatic limit, we can restrict ourselves to non-relativistic dynamics, which are defined by the Hamiltonian
\begin{equation}
\mathcal{H} = \frac{p_0}{\ell  m}+ \frac{1}{\ell ^2 m} + \frac{\alpha}{|\mathbf{x}|}\,,\qquad \alpha=qC\,, \label{Hdef1}
\end{equation}
and the Poisson brackets
\begin{equation}
\{x^i, x^j\} = 2\ell  \,\varepsilon^{ijk} x^k, \qquad \{x^i, p_j\} =\ell p_0 \,\delta^i_{j}-\ell \,\varepsilon^{ikj}\,p_k\,,\qquad \{p_i,p_j\}=0\,.
\end{equation}
The dynamical variables are assumed to be subject to the constraint $p_0^2+\mathbf{p}^2=1/\ell^2$. 
In the standard notation of vector algebra, the Hamiltonian equations of motion take the form
\begin{equation} \label{ueq1}
    \dot p_0=- \alpha \ell \,\frac{\mathbf{p}\cdot\mathbf{x}}{\;|\mathbf{x}|^3}\,, \qquad \dot{\mathbf{p}}=\alpha \ell  \, p_0\,\frac{\mathbf{x}}{\;|\mathbf{x}|^3}+\alpha \ell\, \frac{\mathbf{p}\times \mathbf{x}}{\;|\mathbf{x}|^3}\,, \qquad m\dot{\mathbf{x}}=\mathbf{p}\,.
\end{equation}
In the commutative limit when $\ell\rightarrow 0$ and  $\ell p_0\rightarrow 1$,  we arrive at the familiar equations of the Kepler problem. {In different coordinates, 
the Hamiltonian equations (\ref{ueq1}) have been  integrated in \cite{A2,A3}.}

\subsection{Conservation laws
}
As explained in the previous section, the Hamiltonian equations  (\ref{ueq1}) enjoy a number of conservation lows that make them integrable. 
First of all, the total energy defined by the Hamiltonian (\ref{Hdef1}) is conserved. For the repulsive Coulomb potential ($\alpha >0$), the energy is always non-negative, $\mathcal{H}\geq 0$. In the case of attraction ($\alpha<0$), the energy is bounded above by the `critical energy' $E_c=2/\ell^2m$, which coincides with the maximal value of the kinetic energy of the particle.  Geometrically, it corresponds to the northern pole of the momentum $3$-sphere.  The upper bound $\mathcal{H}\leq E_{c}$ obviously disappears in the commutative limit $\ell\rightarrow 0$. 

The rotational invariance of the system gives rise to one more conserved quantity, namely, the deformed angular momentum vector
\begin{equation}
\mathbf{L} = \ell p_0(\mathbf{x}\times \mathbf{p})\,+ \ell  \, \mathbf{p} \times (\mathbf{x}\times \mathbf{p}) \,.
\label{Ldeform1}
\end{equation}
By construction,
\begin{equation}
\dot{\mathbf{L}}=\{\mathbf{L},\mathcal{H}\}=0\qquad \mbox{and}\qquad \lim_{\ell \to 0} \mathbf{L} = \mathbf{x} \times \mathbf{p}\,. 
\end{equation}
Interestingly enough the square of the deformed angular momentum does not depend on the deformation parameter,
\begin{equation}
L^2:=|\mathbf{L}|^2  = |\mathbf{x}\times \mathbf{p}|^2.
\end{equation}

 Besides the angular momentum, the system admits an additional conserved vector of the form {\cite{A2} }
\begin{equation}
\mathbf{A} = \mathbf{p} \times \mathbf{L} + \frac{m\alpha}{|\mathbf{x}|}\,\big(\mathbf{x} - \ell \, \mathbf{L}\big) \,.\label{RLvec1}
\end{equation}
One can check that $\dot{\mathbf{A}}=\{\mathbf{A},\mathcal{H}\}=0$. Clearly, $\mathbf{A}$ goes into the standard Laplace--Runge--Lenz vector as $\ell  \to 0$. The square of the deformed Laplace--Runge--Lenz vector $\mathbf{A}$ is given by
\begin{equation}
A^2 = \alpha^2 m^2 + 2m\,L^2\, \mathcal{H} \big(1- \mathcal{H}/E_c \big)\,. 
\label{AA1}
\end{equation}
{Since $A^2\geq 0$, we get a nontrivial inequality 
\begin{equation}
    L^2 \leq \frac{\alpha^2 m}{2 | \mathcal{H} \big(1-\mathcal{H}/E_c  \big) |}
    \end{equation}
provided that $\alpha >0$ and $\mathcal{H}>E_c$ or $\alpha <0$ and $\mathcal{H}<0$. It sets an upper bound on possible values of the angular momentum. 
}

The conserved vector quantities close the following algebra w.r.t. the Poisson brackets {\cite{A2}}: 
\begin{eqnarray}
\{L_i, L_j\} &=& \varepsilon_{ijk} L_k\,, \qquad \{A_i, L_j\} = \varepsilon_{ijk} A_k\,, \nonumber\\[3mm]
\qquad \{A_i, A_j\} &=& -2m  \mathcal{H}\big(1- \mathcal{H}/E_c\big)\,\varepsilon_{ijk} L_k\,. 
\end{eqnarray}
In the commutative limit $E_c\rightarrow \infty$ and we arrive at the familiar Poisson bracket relations of the Kepler problem. 

\subsection{Space trajectories}\label{ST}
With all integrals of motion above we can examine particle's trajectories for various initial data.  Multiplying~\eqref{Ldeform1} by $\mathbf{x}$, we obtain
\begin{equation}
\mathbf{L} \cdot \mathbf{x} = \ell  L^2\,.
\end{equation}
Therefore, all space trajectories are plane curves.  The corresponding planes are orthogonal to the deformed angular momentum $\mathbf{L}$ and do not pass through the origin unless $\mathbf{L}=\mathbf{0}$. The distance $d$ between the origin and the trajectory plane is equal to 
\begin{equation}
d = \ell \, L.
\end{equation}
It is convenient to introduce the shifted position vector
\begin{equation}
\mathbf{R} = \mathbf{x} - \ell \, \mathbf{L}.
\end{equation}
As in the commutative case, both the deformed Laplace--Runge--Lenz vector~\eqref{RLvec1} and $\mathbf{R}$ lie in the orbit plane, 
\begin{equation}
\mathbf{L} \cdot \mathbf{A} = 0\,,\qquad\mathbf{L} \cdot \mathbf{R} = 0\,.
\end{equation}
Therefore we can parametrize the orbit  by the length  $R = |\mathbf{R}|$ and the angle $\theta$ between the vectors $\mathbf{A}$ and $\mathbf{R}$. It follows from the definition
\eqref{RLvec1} that
\begin{equation}
\mathbf{A}\cdot\mathbf{R} = \mathbf{R}\cdot (\mathbf{p}\times \mathbf{L}) + \frac{\alpha m\,R^2}{|\mathbf{x}|}.
 \end{equation}
Making a circular shift in the scalar triple product and using the identity
\begin{equation}
   \mathbf{R} \times \mathbf{p} = \ell p_0\,\mathbf{L} \,,
   \label{RDarb1}
\end{equation}
we can write
\begin{equation}
 \mathbf{R}\cdot (\mathbf{p}\times \mathbf{L}) =  \mathbf{L}\cdot (\mathbf{R} \times \mathbf{p}) \,,
 =   \ell p_0\,L^2,
\end{equation}
whence
\begin{equation}
A\, R\, \cos\,\theta =  \ell p_0\,L^2 + \frac{\alpha m\,R^2}{|\mathbf{x}|}. \label{interme1}
\end{equation}
Since the energy is conserved, $
\mathcal{H} = E=\mathrm{const}$, 
the expression~\eqref{Hdef1} for the Hamiltonian implies that
\begin{equation}
\ell p_0 = 1-E\,\ell ^2 m + \frac{\alpha\,\ell ^2 m}{|\mathbf{x}|}\,.
\end{equation}
Rel.~\eqref{interme1} can now be rewritten as 
\begin{equation}
A\, R\, \cos{\,\theta} = \big(1-2E/E_c\big)\,L^2 + \frac{m\,\alpha}{|\mathbf{x}|}\,\big(R^2 + \ell ^2 L^2\big).
\end{equation} 
Finally, noticing that $
|\mathbf{x}|^2={R^2 + \ell ^2 L^2}$,
we arrive at the orbit equation 
\begin{equation} \label{traje1}
A\, R\, \cos{\,\theta} = \big(1-2E/E_c\big)\,L^2 + m\,\alpha\, \sqrt{R^2 + \ell ^2 L^2},
\end{equation}
where the constant $A$ is expressed  through the integrals of motion $L$ and $E$ as
\begin{equation}
A= \sqrt{\alpha^2 m^2 + 2m\,L^2\, E\big(1 - E/E_c \big) }\,,
\end{equation}
c.f. Eq.~\eqref{AA1}.  

Let us take a closer look at the particle's trajectories. By introducing the Cartesian coordinates
in the plane of the orbit,
\begin{equation}
X= R\, \cos{\theta}\,,\qquad Y=R\, \sin{\theta},
\end{equation} 
we can rewrite Eq.~\eqref{traje1} as follows:
\begin{equation} \label{trajeBis1}
A\,X  - \big(1-2E/E_c\big)\,L^2 = m\,\alpha\, \sqrt{(X^2+Y^2) + \ell ^2 L^2}\,.
\end{equation}

\paragraph{The case of $E\neq E_c$ and $E\neq 0$.}
First, consider the situation where $L\neq 0$.
By squaring both sides of Eq.~\eqref{trajeBis1}, after a few simplifications which involve Eq.~\eqref{AA1},
we get  the equation
\begin{equation}
\omega\,(X-X_0)^2  + \xi \,Y^2 = 1\,,
\end{equation}
where
\begin{equation}
\omega = \frac{4 E^2 (1-E/E_c)^2}{\alpha^2}\,, \qquad \xi = -\frac{2m\, E\, (1-E/E_c)}{L^2}\,, \qquad X_0 = \frac{(1-2E/E_c ) A}{2 m\,E\, (1-  E/E_c)}\,.
\end{equation}  
Therefore we have elliptic motion iff $\xi>0$, that is,
\begin{equation}
E < 0 \quad \mbox{or}\quad  E  > E_c\,.
\end{equation}
The motion is hyperbolic iff $\xi<0$ or, what is the same,
\begin{equation}
0 < E< E_c\,.
\end{equation}

\begin{figure}[t]
\center{{\includegraphics[width=0.8\linewidth]{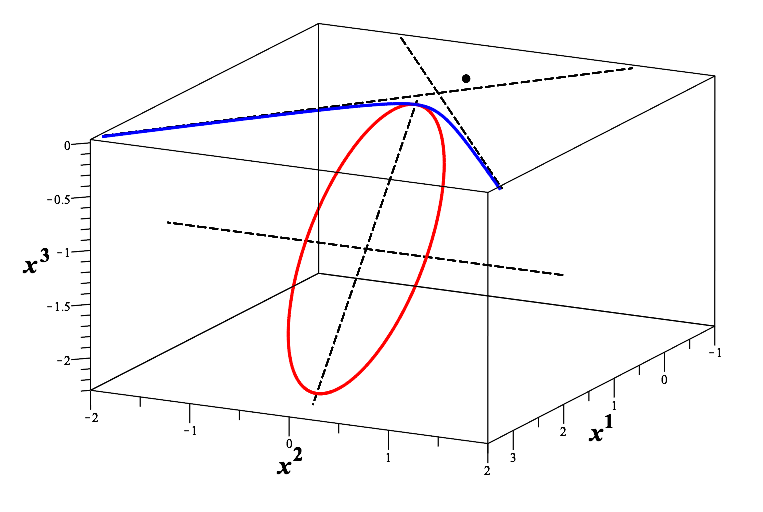}}}  \\
\caption{{{\sl Confinement due to noncommutativity for the repulsive potential with $\alpha=-1$.}  The blue curve is a  hyperbolic trajectory for $\ell=0$. The red curve is an elliptic trajectory for $\ell = 3/2$.  In both cases $m=1$ and the initial conditions are chosen as follows: $x^1(0)=1$, $x^2(0)=x^3(0)=0$, $p_1(0)=p_3(0)=0$, $p_2(0) =1/2$. In the noncommutative case ($\ell=3/2$), the initial value of $p_0$ is chosen in the {southern} hemisphere.}}
 \label{fig1}
\end{figure}

For elliptic motion, the semi-axes of the ellipse are equal to
$1/\sqrt{\omega}$ and $1/\sqrt{\xi}$, respectively. In the limit $L\to 0$,
the minor semi-axis $1/\sqrt{\xi}$ vanishes and the motion becomes radial along the line segment 
\begin{equation}
|X-X_0| \leq 1/\sqrt{\omega}\,,\qquad Y=0\,.   
\end{equation}
The first novelty compared to the commutative case is the possibility of bounded (elliptic or radial) motion for the \emph{repulsive} potential ($\alpha>0$).
In this case, the energy region $E>E_c$ is perfectly accessible when the particle is sufficiently close to the centre. Applying Eqs.~\eqref{CE} and \eqref{Cdef} with $T_{max}=E_c$ and $T_{min}=0$, we arrive at the following conclusion: Any trajectory
that passes through the `trapping region'
\begin{equation}
\mathcal{C} =    
\{\mathbf{x} \in \mathbb{R}^3\; |\; |\mathbf{x}| < \alpha/E_c \} \,,
\end{equation}
remains confined in the compact region 
\begin{equation}
\mathcal{C}_E =  
\big\{ \mathbf{x}\in \mathbb{R}^3\;\big|\;|\;\mathbf{x}| \leq {\alpha}/{(E-E_c)} \big\}\, .
\end{equation}  
{This effect is illustrated in Fig.~\ref{fig1}.}

When the motion is hyperbolic, the semi-axes of the hyperbola are equal to
$1/\sqrt{\omega}$ and $1/\sqrt{|\xi|}$.   Two branches of the hyperbola are located in the 
regions 
\begin{equation}
X \leq \frac{A}{m\,(E-E_c) } + \frac{E_c\,(m|\alpha|-A)}{2\,m \,E\, (E-E_c)} \label{branch1}
\end{equation}
and 
\begin{equation}
X\geq \frac{A}{m\, E} -  \frac{E_c\,(m|\alpha|-A)}{2\,m \,E\, (E-E_c)}\,, \label{branch2}
\end{equation}
where the boundary values of $X$  correspond to the intersection points with the $X$-axis. These two branches correspond to $\alpha<0$ and $\alpha>0$ respectively\footnote{This statement is obvious at $\ell= 0$. For $\ell >0$, it holds by continuity. }.

\begin{figure}[t]
\center{{\includegraphics[width=0.8\linewidth]{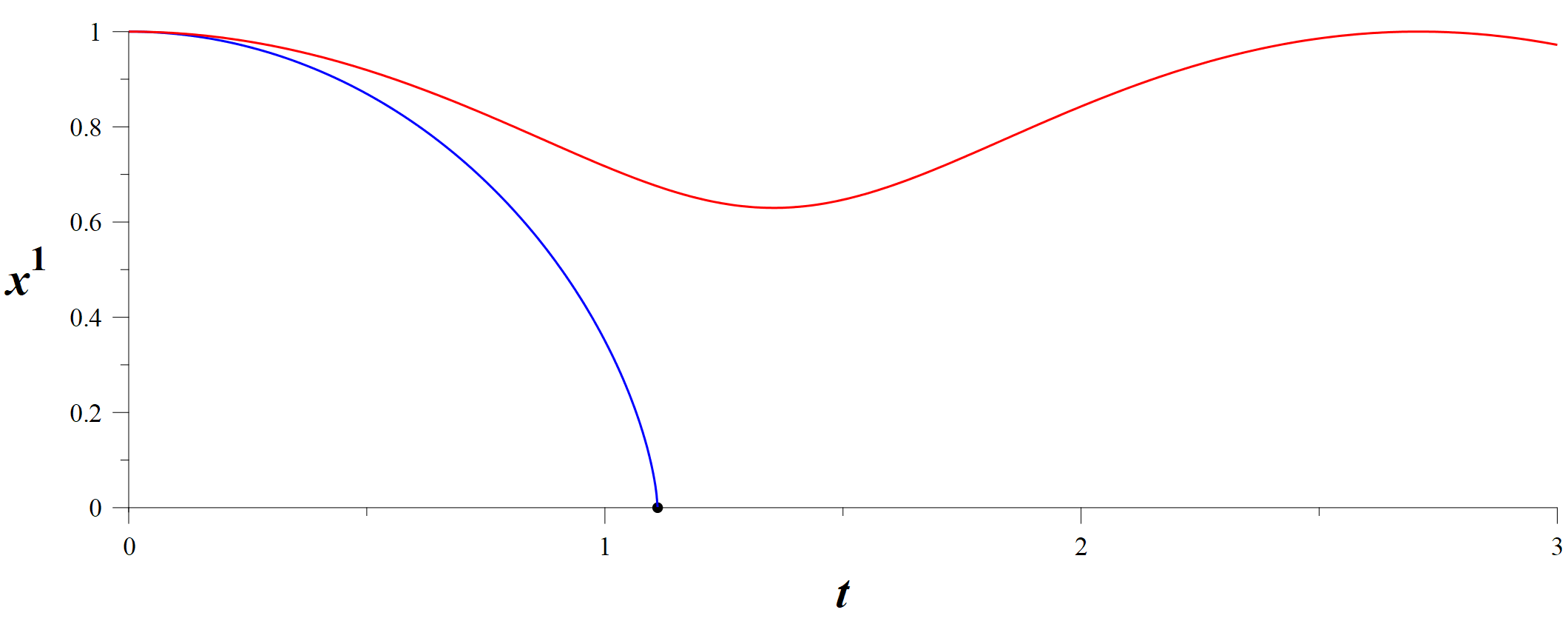}}}  \\
\caption{{{\sl Radial motion along the axis $x^1$ for the attractive potential with $\alpha=1$. }  The blue line represents a solution to the equations of motion with $\ell=0$. In a finite time $t\approx 1.1$, the particle falls on the centre. The red line represents a solution for $\ell=3/2$. Though the potential is attractive, the particle cannot reach the centre, oscillating along the $x^1$ axis.   In both cases $m=1$ and the initial conditions are chosen as follows: $x^1(0)=1$, $x^2(0)=x^3(0)=0$, $p_1(0)=p_2(0)=p_3(0)=0$. In the noncommutative case ($\ell=3/2$), the initial value of $p_0$ is chosen in the southern hemisphere.
}}
 \label{fig2}
\end{figure}

In the limit $L\to 0$,  the trajectories `shrink' to the rays
\begin{equation} 
Y=0,\qquad X \leq \frac{|\alpha|}{m\,(E-E_c) }<0 \qquad\mbox{for}\qquad \alpha<0 \label{nofallcenter1}
\end{equation}
and
\begin{equation} 
Y=0,\qquad X \geq \frac{|\alpha|}{m\,E }>0\qquad\mbox{for}\qquad \alpha>0\,.
\label{ray21} 
\end{equation}
Eq.~\eqref{nofallcenter1} describes another important novelty: even though the potential is attractive, the particle cannot fall to the centre! Indeed, the region $|X| <|\alpha|/(E_c -E)$ is forbidden, in agreement with Eq.~\eqref{nofallregion}. A numerical solution that illustrates this effect is presented in Fig.~\ref{fig2}.

\paragraph{The case of $E=0$.}
One can easily see from Eq.~\eqref{traje1} that, whenever $E=0$ and $L\neq 0$, the motion is parabolic. The  trajectories
are given by the plane curves
\begin{equation}
Y^2 = \nu \, (X-\zeta)   
\label{parabolic11}
\end{equation}
with
\begin{equation}
 \nu =   \frac{2\,L^2}{\alpha\,m},\qquad \zeta = \frac{\alpha^2\ell ^2 m^2-L^2}{2\,\alpha\,m}\,. 
 \label{parabolic21}
\end{equation}
The equality $E=0$ is only possible for the \emph{attractive} potential, i.e., $\alpha<0$.
Therefore, the parabola is located in the region $X\leq \zeta$. In the limit $L\to 0$, the parabola `shrinks' to the ray 
\begin{equation}
X\leq \alpha/E_c <0\,, \qquad Y =0
\label{ray1}
\end{equation}
and the motion becomes radial. Of course, this formula can also be obtained from Eq.~\eqref{nofallcenter1} by passing to the limit $E\to 0$. Despite the fact that the potential is attractive, the particle with $L=0$ cannot fall to the centre: the region $|x| < |\alpha|/E_c $ is forbidden.

\paragraph{The case of $E= E_c$. } For $E=E_c$ and $L\neq 0$, the motion is also parabolic. Indeed,   relation~\eqref{traje1} yields  equations~\eqref{parabolic11} and~\eqref{parabolic21}. However, the equality $E=E_c$ is only possible for $\alpha>0$. Therefore, in contrast to the previous case, 
our parabola is located in the region $X\geq \zeta$. The possibility of having the parabolic motion for the repulsive potential is the third nontrivial property of the model under consideration. As $L$ goes to zero this parabola `shrinks' to the ray
\begin{equation}
X\geq {\alpha}/{E_c},
\end{equation}
so the motion becomes radial.
This result perfectly matches our previous conclusion~\eqref{ray21} when $E\to E_c$.

{
To summarize, all space trajectories resulting from the Hamiltonian  equations (\ref{ueq1}) are given by conic sections, as in the ordinary Kepler problem.  Moreover, for a given kinetic energy, the Coulomb potential is proved to be the only central power-law potential with such a property \cite{A3}. }

\subsection{Other Hamiltonians}

\begin{figure}[t]
\center{{\includegraphics[width=0.8\linewidth]{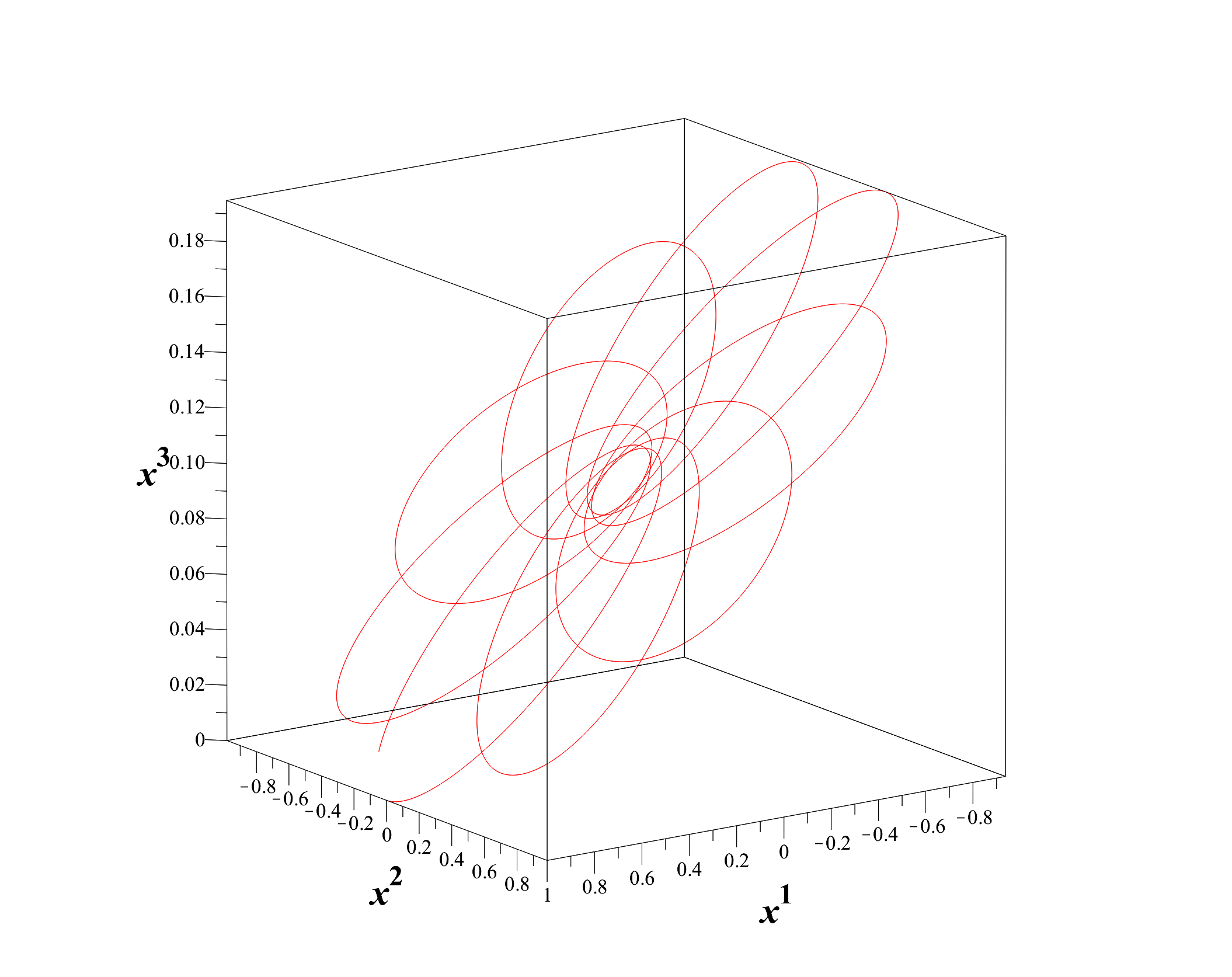}}}  \\
\caption{{{\sl A typical trajectory for the alternative choice~\eqref{HdefAltern} of the kinetic energy.}  The parameters are chosen as follows: $\alpha=-1$, $\ell =1/5$, $m=1$. The initial conditions are given by $x^1(0)=1$, $x^2(0)=x^3(0)=0$, $p_1(0)=p_3(0)=0$, $p_2(0) =1/2$. For these parameters, the whole trajectory projects onto the southern hemisphere of the momenta $3$-sphere.}}
 \label{fig3}
\end{figure}

At this point, the reader may wonder what the advantage is of the Hamiltonian (\ref{Hdef1}) over other options. A partial answer to this question is that the above choice for the kinetic energy function results in a superintegrable system. It is the presence of an additional integral of motion (\ref{RLvec1}) that significantly simplifies the analysis of trajectories. An arbitrary choice of kinetic energy will most likely not provide an additional conserved quantity. For example, the simplest candidate consistent with the commutative limit would be

\begin{equation}
\mathcal{H} = \frac{|\mathbf{p}|^2}{2 m} + \frac{\alpha}{|\mathbf{x}|}\,,\qquad \alpha=qC\,. \label{HdefAltern}
\end{equation}
All our conclusions regarding the confinement for the repulsive potential and the impossibility to fall into the center for the attractive potential  remain valid. Moreover, since the rotational invariance is still there, the (deformed) angular momentum is conserved, and the system is integrable. 
The numerical analysis of the Hamiltonian equations
\begin{equation}
\dot{x}^i= \{x^i, \mathcal{H}\}, \qquad \dot{p}_i= \{p_i, \mathcal{H}\} 
\end{equation}
shows that the orbits, still being flat curves,  exhibit a precession as in Fig.~\ref{fig3}. Therefore, we do not expect to find any reasonable deformation of the Laplace--Runge--Lenz vector which would yield superintegrability. A detailed analysis of this  Hamiltonian is beyond the scope of this paper.

\section*{Acknowledgments} 

V.G.K. acknowledges support from the National Council for Scientific and Technological Development (CNPq) Grant 304130/2021-4.

\end{document}